\documentclass[%
 aip,
 amsmath,amssymb,
preprint,                       
]{revtex4-1}

\usepackage[dvipdfmx]{graphicx}
\usepackage{dcolumn}
\usepackage{bm}

\usepackage[utf8]{inputenc}
\usepackage[T1]{fontenc}
\usepackage{mathptmx}
\usepackage{etoolbox}
\usepackage{xcolor}
\usepackage{ulem}

\newcommand {\SRO}{SrRuO$_3$}
\newcommand {\STO}{SrTiO$_3$}
\newcommand {\deltarho}{$\Delta \rho_{xx}$}

\makeatletter
\def\@email#1#2{%
 \endgroup
 \patchcmd{\titleblock@produce}
  {\frontmatter@RRAPformat}
  {\frontmatter@RRAPformat{\produce@RRAP{*#1\href{mailto:#2}{#2}}}\frontmatter@RRAPformat}
  {}{}
}%
\makeatother
\begin{document}

\preprint{AIP/123-QED}

\title{Unconventional two-dimensional quantum oscillations in three-dimensional thick SrRuO$_3$ films}
\author{Yuta Matsuki}
\affiliation{ 
Department of Physics, Tokyo Institute of Technology, Tokyo, 152-8551, Japan
}
\author{Shinichi Nishihaya}%
\affiliation{ 
Department of Physics, Tokyo Institute of Technology, Tokyo, 152-8551, Japan
}

\author{Markus Kriener}
\affiliation{ 
RIKEN Center for Emergent Matter Science (CEMS), Wako 351-0198, Japan
}%

\author{Ren Oshima}
\affiliation{ 
Department of Physics, Tokyo Institute of Technology, Tokyo, 152-8551, Japan
}

\author{Fumiya Miwa}
\affiliation{ 
Department of Physics, Tokyo Institute of Technology, Tokyo, 152-8551, Japan
}

\author{Masaki Uchida}
\altaffiliation{Electronic mail: m.uchida@phys.titech.ac.jp}
\affiliation{ 
Department of Physics, Tokyo Institute of Technology, Tokyo, 152-8551, Japan
}


\begin{abstract}
\SRO\ is a prototypical transition metal oxide which hosts rich physical properties including itinerant ferromagnetism, high conductivity, and intrinsic Hall effect originating in the Weyl points. Recently, high-quality \SRO\ films with residual resistivity ratios of more than 50 have been reported to exhibit quantum oscillations at low temperatures in spite of its strong electron correlation. While the origin of the oscillations has been discussed in relation to Weyl orbits based on the Weyl semimetal band structure, so far experimentally reported results are neither consistent with each other nor with theoretically expected behavior, leaving the origin of the oscillations in \SRO\ films still elusive. In this report, we have carefully evaluated the quantum oscillations observed in three-dimensional thick \SRO\ films with a high residual resistivity ratio of RRR = 82. We reveal the coexistence of two oscillation components both derived from two-dimensional electronic states and with slightly different masses, suggesting the involvement of the surface Fermi arc states formed between different Weyl point pairs.
\end{abstract}

\maketitle


 In recent condensed matter physics, topological phases have attracted growing attention owing to their unique conducting states originating in topological band structures. Especially in topological semimetals characterized by Weyl points, three-dimensional (3D) band degenerate points in the momentum space, Fermi arc states emerge connecting the Weyl points projected onto the top and bottom surfaces. Under magnetic field, a unique cyclotron orbit called Weyl orbit is expected to be formed, where electrons loop between Fermi arc states on the top and bottom surfaces through bulk tunneling\cite{SdH_semimetal_calc}. Experimentally, Shubnikov-de Haas (SdH) oscillations associated with Weyl orbits have been reported in several topological semimetals including Dirac semimetal Cd$_3$As$_2$\cite{Cd3As2_2016,Cd3As2_QH,Cd3As2_gate,Cd3As2_1,Cd3As2_2} and inversion-symmetry broken Weyl semimetals WTe$_2$\cite{WTe2} and NbAs\cite{NbAs}. 

 
 Among Dirac and Weyl semimetals, \SRO\ is known as a prototypical material\cite{review} where the observation of an intrinsic anomalous Hall effect originating in magnetic monopoles has been reported for the first time\cite{First_AHE}. Later \SRO\ has been predicted and demonstrated to be a time-reversal-symmetry (TRS) broken Weyl semimetal \cite{SRO_Weyl_theory}. Recently, high-quality \SRO\ thin films with residual resistivity ratios (RRR) exceeding 50 have been reported to exhibit SdH oscillations despite its strong electron correlation\cite{Mackenzie,SRO_multi_SdH1,SRO_multi_SdH2,NTT_dimensionality,ACS_dimensionality}. This has attracted interest as the TRS broken Weyl semimetal possibly exhibiting the Weyl orbit\cite{SRO_multi_SdH1,SRO_multi_SdH2}, which would allow access to the rich physics of Weyl orbits coupled with magnetism. However, the origin of the SdH oscillations in \SRO\ films is still elusive due to its inconsistent observations between different groups\cite{SRO_multi_SdH1,SRO_multi_SdH2,NTT_dimensionality,ACS_dimensionality}. Since the Weyl orbit consists of a two-dimensional (2D) Fermi pocket surrounded by the two surface Fermi arcs and the bulk 3D Fermi pockets, surface and bulk oscillation components should appear obeying the frequency relation $F_{\textrm{bulk}}$ < $F_{\textrm{surface}}$ and effective mass relation $m_{\textrm{bulk}}^*$ < $m_{\textrm{surface}}^*$. This feature of the Weyl orbit is important also for distinguishing it from other trivial mechanisms including 2D electronic states as induced by quantum confinement, where the occupied subbands give multiple 2D oscillation components with similar masses ($F_{\textrm{sub}1}$ > $F_{\textrm{sub}2}$, $m_{\textrm{sub}1}^*$ $\sim$ $m_{\textrm{sub}2}^*$). Table I summarizes quantum-oscillation features reported for \SRO\ films from different groups, in comparison with representative theoritical models. Previous experimental reports of the SdH oscillations are neither consistent with each other, especially in terms of the dimensionality of the oscillation components, nor with the theoretical models. The motivation of this study is to carefully examine dimensionalities and mass values of multiple oscillation components in \SRO\ films. By measuring high-quality and three-dimensional thick \SRO\ films, we have identified that the oscillation components are derived from two different 2D electronic states obeying the mass relation $m_1^*$ > $m_2^*$.


 \SRO\ films were grown in an oxide molecular beam epitaxy system equipped with laser heating. This setup has been also used to fabricate high-quality epitaxial films of other Sr$_{n+1}$Ru$_n$O$_{3n+1}$ Ruddlesden Popper series including Sr$_2$RuO$_4$ ($T_{\textrm{c,mid}}$ = 1.2 K)\cite{Sr2RuO4_APL,Sr2RuO4_upper,Sr2RuO4_PRB} and Sr$_3$Ru$_2$O$_7$ (RRR = 24)\cite{Oshima}. \SRO\ films were grown on \STO\ (001) substrate, which has only a small lattice mismatch (-0.6\%) with respect to pseudocubic SrRuO$_3$. Prior to the growth, \STO\ substrates were annealed at 870$^{\circ}$C with flowing O$_3$ (20\%)+O$_2$ (80\%) mixture gas inside the MBE chamber to achieve a smoother surface. 
\SRO\ films were grown by co-evaporation of elemental sources; 4N Sr from a conventional Knudsen cell, 3N5 Ru from an electron beam evaporator, and the O$_3$(20\%)+O$_2$(80\%) mixture gas from a Meidensya ozone generator. The film thickness was designed to be more than 60 nm in order to suppress possible impurity diffusion from the substrate. Due to the highly volatile nature of Ru compared to Sr, the growth was performed under Ru-rich conditions. 


Figure 1(a) presents the orthorhombic structure of \SRO\ \cite{ortho_1,ortho_2} with a dominant type of domains (domain A) formed on the \STO\ substrate, depending on the tilting direction of the [110]$_{\textrm o}$ axis with respect to the [100]$_{\textrm c}$ substrate miscut direction. The subscripts ``o'' and ``c'' denote orthorhombic and cubic, respectively. The [110]$_{\textrm o}$ axis is slightly tilted toward the [$\bar{1}$00]$_{\textrm c}$ direction in domain A and it is rotated clockwise by every 90$^\circ$ in domains B, C, and D. Figure 1(b) shows the azimuthal $\phi$ scan obtained for the (021)$_{\textrm o}$ Bragg peak of a \SRO\ film with a thickness of $t$ = 72.4 nm. The intensity of domain A is indeed 10 times larger than that of any other domains. Figure 1(c) exhibits temperature dependence of the longitudinal resistivity measured on this \SRO\ film. It exhibits metallic behavior with a ferromagnetic transition temperature of $T_{\textrm C}$ = 152 K, which is slightly lower than the bulk value of 160 K due to compressive strain (-0.6\%) from the \STO\ substrate\cite{decrease_Tc_compressive}. Its RRR value of 82 is comparable to the values previously reported for high-quality \SRO\ films\cite{SRO_multi_SdH1,SRO_multi_SdH2,topRRR,Schlom,Izumi,Mackenzie}. Figure 1(d) presents magnetoresistance MR = ($\rho(B)$-$\rho$(0 T))/$\rho$(0 T) taken at $T$ = 2 K. This exhibits a large linear positive MR of 35 \% at 9 T, suggesting high mobility carriers in the \SRO\ film. The magnification of MR shows a small hysteresis loop originating in the ferromagnetic ground state.


Figure 2 summarizes analysis of SdH oscillations observed in the \SRO\ film. The oscillatory component $\Delta\rho_{xx}$ is extracted by subtracting a polynomial background from $\rho_{xx}$. \textcolor{black}{SdH oscillations are more clearly observed in films with higher RRR due to suppression of the scattering by impurities and at crystalline domain boundaries.} As shown in the FFT spectra of $\Delta\rho_{xx}$ in Fig. 2(b), there are two different oscillation components $F_1$ and $F_2$, as also reported in previous studies\cite{SRO_multi_SdH1,SRO_multi_SdH2}. For the mass analysis of such multi-component oscillations, in general it is necessary to apply the Lifshitz-Kosevich formula to the FFT amplitude obtained by extracting the $\Delta\rho_{xx}$ curve with a certain window function centered at $(1/B)_{\textrm c}$ and with the width $\Delta(1/B)$\cite{Accurate_mass}. The mass can be extracted from the following formula
\begin{equation}
\frac{A(T)}{T} = \frac{A_0}{\sinh(2\pi k_{\textrm B}T/\hbar\omega_{\textrm c})},
\label{mass_fit}
\end{equation}
where $A(T)$ is the FFT amplitude at each temperature and $1/\omega_{\textrm c}$ = $(m^*/e)\Delta(1/B)$ is the cyclotron frequency. Importantly, the FFT amplitude sensitively depends on the choice of the window type and the parameters $(1/B)_{\textrm c}$ and $\Delta(1/B)$\cite{Accurate_mass}. In particular, choosing a too large $\Delta(1/B)$ gives rise to a significant overestimation of $A(T)$ and underestimation of the effective mass value, while a too small $\Delta(1/B)$ yields an artificial oscillatory behavior of $A(T)$ with respect to $\Delta(1/B)$ due to an insufficient number of oscillations in the FFT range. Taking into account the frequencies $F_1$ and $F_2$ and their appearance above 3 T, we employ a rectangular window function with fixing $\Delta(1/B)$ = 0.17 T$^{-1}$ and varying $(1/B)_{\textrm c}$ from 0.1575 to 0.2100 T$^{-1}$. Figure 2(b) presents the FFT spectra obtained with $(1/B)_{\textrm c}$ = 0.185 T$^{-1}$, where two oscillation components distinctly appear at $F_1$ = 27 T and $F_2$ = 37 T. At higher temperatures, their oscillation amplitudes become smaller and it gets more difficult to distinguish the two peaks without fitting. So here we use Gaussian fitting in order to estimate the $F_1$ and $F_2$ amplitudes.
Figure 2(c) shows temperature dependence of the FFT amplitudes for the case of $(1/B)_{\textrm c}$ = 0.185 T$^{-1}$, and Figure 2(d) shows $(1/B)_{\textrm c}$ dependence of the extracted effective mass values. $m_1^*$ is almost constant around 0.32$m_0$ for the higher field range (0.1575 < $(1/B)_{\textrm c}$ < 0.185 T$^{-1}$), while $m_2^*$ is constant around 0.24$m_0$ for the lower field range (0.185 < $(1/B)_{\textrm c}$ < 0.21 T$^{-1}$). This ensures the validity of the present mass analysis adopting $(1/B)_{\textrm c}$ = 0.185 T$^{-1}$. Importantly, the relation $m_1\*$ > $m_2\*$ is obtained regardless of the $(1/B)_{\textrm c}$ value, indicating that an electronic state giving the $F_2$ component has a larger Fermi surface with a smaller band mass as compared to the one giving the $F_1$ component.


Next we examine dimensionalities of the oscillation components. Figure 3(a) plots field angle dependence of the oscillations as a function of 1/$B$ at $T$ = 2 K. $\theta$ is the angle measured from the out-of-plane [001]$_{\textrm c}$ towards [010]$_{\textrm c}$. As presented in Figs. 3(b) and 3(c), both of the oscillation frequencies $F_1$ and $F_2$ increase with increase in $\theta$, and their amplitudes gradually decrease and eventually vanish above $\theta$ = 60$^\circ$. This observation clearly indicates that the Fermi surface structure of both $F_1$ and $F_2$ is two dimensional. Oscillation peaks almost obey the $1/\cos\theta$ scaling up to $\theta$ = 20$^\circ$, but deviate from it above $\theta$ = 20$^\circ$ with a clear phase shift. 

 Detailed angular dependence of the $F_1$ FFT amplitude in Fig. 3(b) reveals that there is a significant drop of the amplitude around $\theta$ = 10$^\circ$, which is induced by only slight tilting of magnetic field from the out-of-plane direction. This can be also confirmed in $\Delta\rho_{xx}$ in Fig. 3(a). In a 2D system, suppression of the oscillation amplitude at a specific tilting angle may occur for example by (i) the Yamaji effect\cite{Yamaji_effect} or (ii) the interference effect between different oscillatory components including the so-called spin-zero effect\cite{CsV3Sb5,ZrTe5,spin_zero}. While the Yamaji effect is expected to exhibit a periodic tan$\theta$ dependence of the oscillation amplitude\cite{CsV3Sb5,Yamaji_tan}, the observed angular dependence does not follow it. Thus, the Yamaji effect does not apply in the present case. As for the interference effect, one possibility is that the interference between the $F_1$ and $F_2$ components, which are close in frequency, results in an accidental suppression of the oscillation amplitude in the measured field range. However, the out-of-plane component of the magnetic field at $\theta$ = 10$^\circ$ only decreases by 1.5\% from $\theta$ = 0$^\circ$, and it is unlikely that the relative phase between the $F_1$ and $F_2$ components changes by such a tiny tilting angle. Another possibility is the spin-zero effect, which is derived from the interference between the Zeeman-split Fermi surfaces. In the Lifshitz-Kosevich formula, the oscillation amplitude related to the Zeeman splitting is calculated by $R_{\textrm s} = \cos\left(\frac{\pi}{2}\frac{gm^*}{m_0\cos\theta}\right)$ with the $g$ factor $g$. Namely, the spin-zero effect occurs when $\frac{gm^*}{m_0\cos\theta} = 2n+1$ holds for an integer $n$, accompanied by a relative phase shift of $\pi$ between the Zeeman-split oscillation components. In the present case, such a $\pi$ phase shift of the quantum oscillations is not observed across $\theta$ = 10$^\circ$, excluding the spin-zero effect. To sum up, the significant suppression of the oscillation amplitude at this small tilting angle remains an open question and may present an important issue.
 
 Finally, we would like to discuss the origin of the 2D electronic states which give rise to the $F_1$ and $F_2$ components. First, the thickness of the measured \SRO\ film is 72.4 nm, much larger than the inverse of the Fermi wave number $\lambda_{\textrm F} = 2\pi/k_{\textrm F}$ (22 nm for $F_1$ = 27 T and 19 nm for $F_2$ = 37 T), indicating that the quantum confinement effect is unlikely as the origin of the 2D electronic states. Also in terms of the relation $m_1^*$ > $m_2^*$ for $F_1$ < $F_2$, the confinement scenario is unlikely. On the other hand, our observation does not exclude a Weyl orbit scenario, if we assume that the two 2D states originate in two different Weyl orbits consisting of different Weyl point pairs in SrRuO$_3$. The absence of the oscillation component from a 3D bulk state with even smaller effective mass may suggest that the bulk Fermi pockets involved in the Weyl orbits are too small and that they already reach the quantum limit under the present magnetic fields. \textcolor{black}{Another possibility is that these two different 2D electronic states originate in asymmetry between the top and bottom surfaces. In order to further verify these scenarios, it is desirable to independently control the top or bottom surface state, for example, by electrostatic gating \cite{SRO_gate,SRO_gate_Weyl}.} 


 In summary, we have grown high-quality three-dimensional thick \SRO\ films by molecular beam epitaxy. We have observed the SdH oscillation with two oscillation components in a \SRO\ film with an RRR of 82, and have determined their dimensionalities and effective mass values through a detailed oscillation analysis. Both $F_1$ and $F_2$ components originate in a 2D electronic structure and their effective mass values are $m_1^*$ = 0.32$m_0$ and $m_2^*$ = 0.24$m_0$. The appearance of such 2D electronic states may suggest the realization of several Weyl orbits consisting of different Weyl point pairs in \SRO\ films.



\begin{acknowledgments}
\textcolor{black}{This work was supported by JST FOREST Program Grant No. JPMJFR202N, Japan, by Grant-in-Aids for Scientific Research JP22H04471, JP22H04501, JP22K18967, JP22K20353, JP23K13666, JP24H01614, and JP24H01654 from MEXT, Japan, by The Asahi Glass Foundation, Japan, by Izumi Science and Technology Foundation, Japan, and by STAR Award funded by the Tokyo Tech Fund, Japan.}
\end{acknowledgments}

\section*{Conflicts of Interest}
The authors have no conflicts to disclose.
\section*{Author Contributions}

\textbf{Yuta Matsuki}: Data curation (lead); Formal analysis (equal); Investigation (lead); Methodology (equal); Visualization (lead); Writing - original draft preparation (lead). 
\textbf{Shinichi Nishihaya}: Formal analysis (equal), Investigation (equal), Methodology (equal), Writing-review \& editing (equal).
\textbf{Markus Kriener}: Formal analysis (supporting), Investigation (equal), Writing-review \& editing (equal).
\textbf{Ren Oshima}: Formal analysis (supporting), Investigation (supporting), Writing-review \& editing (equal).
\textbf{Fumiya Miwa}: Formal analysis (supporting), Investigation (supporting), Writing-review \& editing (equal).
\textbf{Masaki Uchida}: Conceptualization (lead); Funding acquisition (lead); Methodology (lead); Project administration (lead); Supervision (lead); Writing - review \& editing (lead).

\section*{Data Availability}
The data that supports the findings of this study are available
from the corresponding author upon reasonable request.



\newpage 


\begin{figure*}
\begin{center}
\includegraphics[width=17cm]{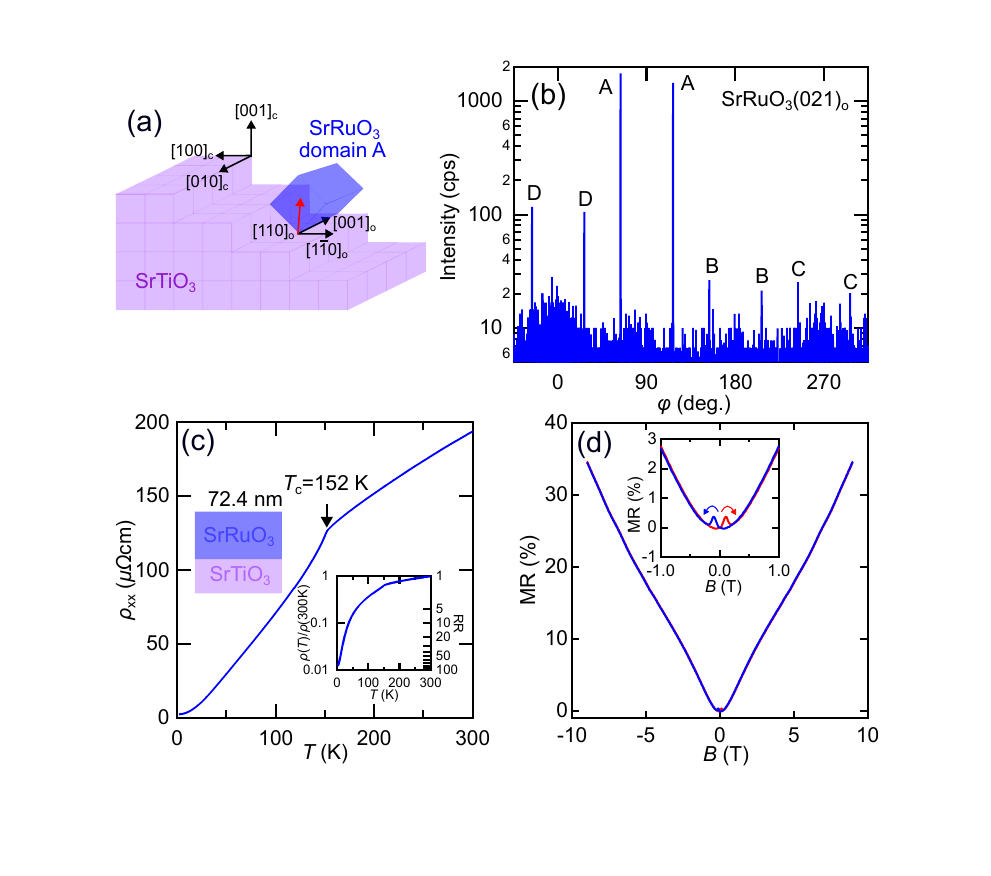}
\caption{
Fundamental structure and transport characterization of a \SRO\ film. (a) The lattice structure of \SRO\ is orthorhombic and it forms four types of domains on the cubic \STO\ substrate depending on in which direction the [110]$_{\textrm o}$ axis is tilted with respect to the [100]$_{\textrm c}$ substrate miscut direction. (b) In-plane x-ray diffraction $\varphi$ scan for (021)$_{\textrm{o}}$ \SRO\ peak.  Domain A, with the [110]$_{\textrm o}$ axis slightly tilted toward [$\bar{1}$00]$_{\textrm c}$, dominantly exists, compared to other domains B-D. (c) Temperature dependence of the longitudinal resistivity $\rho_{xx}$. It exhibits metallic behavior with ferromagnetic transition at $T_{\textrm C}$ = 152 K. The inset shows temperature dependence of $\rho(T)$/$\rho$(300 K) with the resistivity ratio (RR) for the right axis. (d) Magnetoresistance measured at $T$ = 2 K. With sweeping the out-of-plane magnetic field, a linear positive magnetoresistance is observed with MR = ($\rho(B)$-$\rho$(0 T))/$\rho$(0 T) = 35 \% at 9 T. A magnification of the hysteresis around 0 T is shown in the inset.
}
\label{fig1}
\end{center}
\end{figure*}

\begin{table}[h]
    \centering
    \label{Dimensionality}
    \begin{tabular}{cccc} \hline
      & & Fermi surface size (dimensionality) & mass \\ \hline
    Weyl orbit & & $F_{\mathrm{bulk}}$(3D) < $F_{\mathrm{surface}}$(2D) & $m_{\mathrm{bulk}}^*$ < $m_{\mathrm{surface}}^*$ \\  \hline 
    quantum confinement & & $F_{\mathrm{sub1}}$(2D) > $F_{\mathrm{sub2}}$(2D) & $m_{\mathrm{sub1}}^*$ $\sim$ $m_{\mathrm{sub2}}^*$ \\  \hline
    & (Ref. \onlinecite{NTT_dimensionality}) & $F_1$(2D) < $F_2$(3D) & $m_1^*$ < $m_2^*$ \\  \cline{2-4}
    \SRO\ films & (Ref. \onlinecite{ACS_dimensionality}) & $F_1$(2D) < $F_2$(2D) & --- \\  \cline{2-4}
    & (present work) & $F_1$(2D) < $F_2$(2D) & $m_1^*$ > $m_2^*$ \\  \hline 
    \end{tabular}
    \caption{Fermi surface size, dimensionality, and mass relation of multiple quantum oscillations expected in theoretical models and observed for \SRO\ films. $F_1$ and $F_2$ are assigned in decreasing order of frequency.}
\end{table}

\begin{figure}
\begin{center}
\includegraphics[width=17cm]{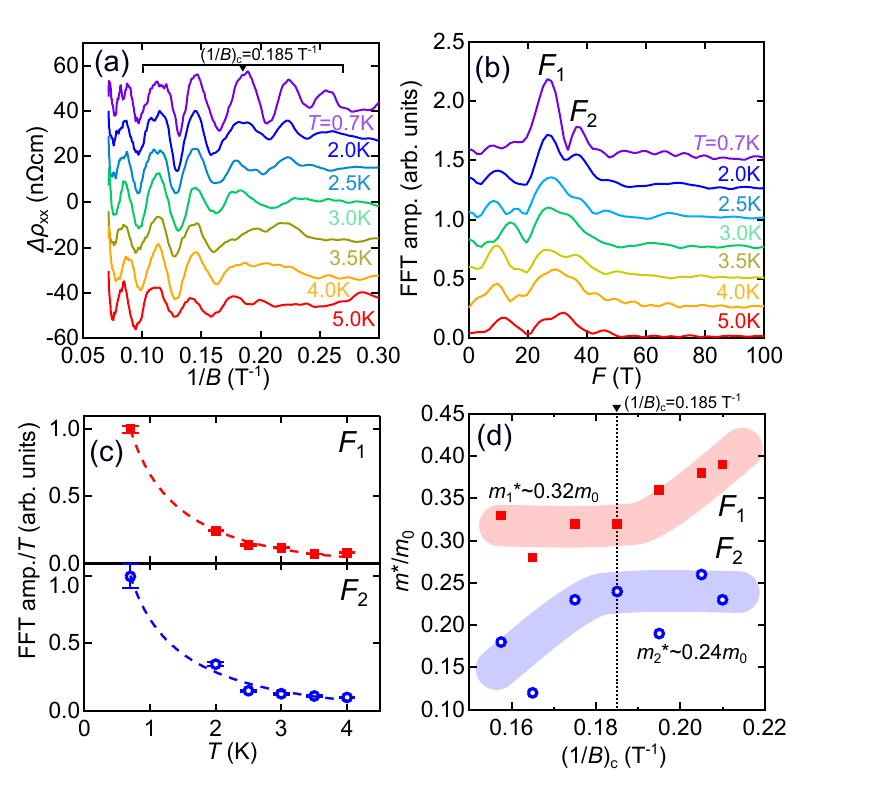}
\caption{
Multiple quantum oscillations in the \SRO\ film. (a) Oscillatory components \deltarho\ taken for the \SRO\ film at $T$ = 0.7-5.0 K, after subtracting a background from $\rho_{xx}$. Non-monotonic behavior with respect to $B$ is observed. (1/$B)_{\textrm c}$ indicates a center value of the inverse of the magnetic field used in the fast Fourier transformation (FFT) analysis. (b) Fourier transformation of \deltarho\ in the field range with (1/$B)_{\rm c}$ = 0.185 T$^{-1}$ in (a). There are two prominent peaks at $F_1$ = 27 T and $F_2$ = 37 T. (c) Temperature dependence of the FFT amplitude in (b) divided by $T$ for $F_1$ and $F_2$, plotted for estimating their effective mass values. The error bar indicates the relative error of the Gaussian fitting. (d) Effective mass values $m_1^*$ and $m_2^*$ obtained from the temperature change in multiple FFT amplitudes, which are systematically estimated using rectangular window functions with various center values (1/$B)_{\textrm c}$ and a fixed width of $\Delta(1/B)$ = 0.17 T$^{-1}$. 
}
\label{fig2}
\end{center}
\end{figure}

\begin{figure*}
\begin{center}
\includegraphics[width=17cm]{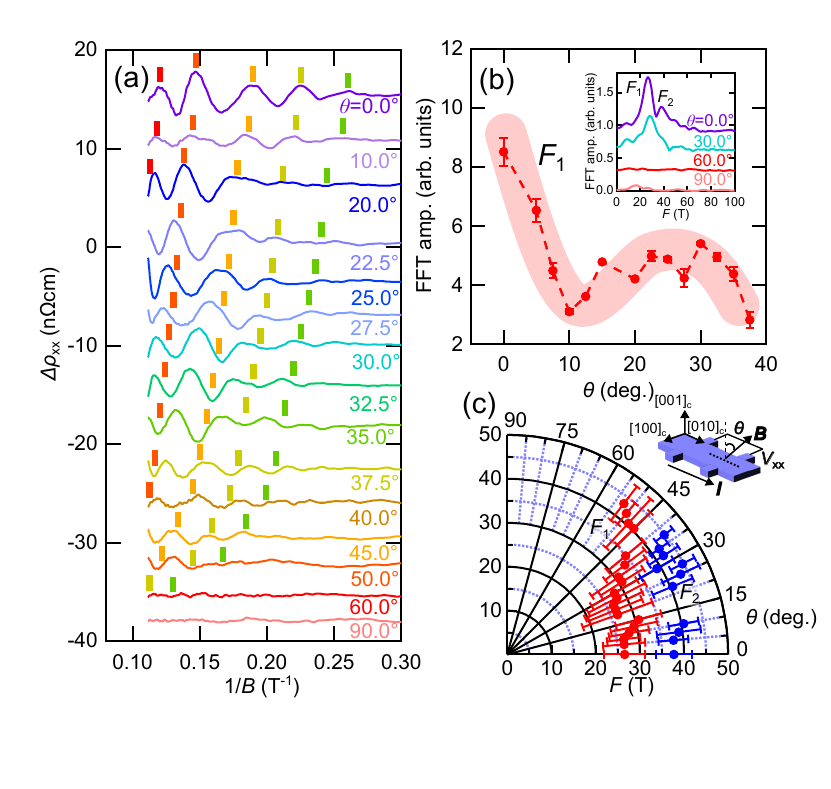}
\caption{
Field-angle dependence of the quantum oscillations. (a) Oscillatory components \deltarho\ taken for various field angles $\theta$ at $T$ = 2 K. $\theta$ is measured from the out-of-plane [001]$_{\textrm c}$ towards the in-plane [010]$_{\textrm c}$ direction. A vertical colored bar signifies the peak positions calculated by applying the simple 1/cos$\theta$ scaling to the $\theta$ = 0$^\circ$ case. (b) Angular dependence of the FFT amplitude of $F_1$. The inset shows the field-angle dependence of the FFT calculated by applying the window function with (1/$B)_{\textrm c}$ = 0.185 T$^{-1}$ and $\Delta(1/B) = 0.17$  T$^{-1}$ for $\theta$ = 0, 30, 60 and 90$^\circ$. (c) Field-angle dependence of the frequencies $F_1$ and $F_2$. A radial bar expresses the FWHM in the Gaussian fitting. A schematic measurement configuration with a Hall bar device is shown in the inset.
}
\label{fig3}
\end{center}
\end{figure*}


\end{document}